\documentclass[nofootinbib,pra,twocolumn]{revtex4}
\usepackage{natbib}
\usepackage{amsfonts}
\usepackage{amsmath}
\usepackage{amssymb}
\usepackage{amsthm}
\usepackage{graphicx}
\usepackage{bm}
\usepackage{bbm}
\usepackage{hyperref}
\newcommand{\be}{\begin{eqnarray} \begin{aligned}}
\newcommand{\ee}{\end{aligned} \end{eqnarray} }
\newcommand{\benn}{\begin{eqnarray*} \begin{aligned}}
\newcommand{\eenn}{\end{aligned} \end{eqnarray*} }

\newcommand{\ran}{\rangle}
\newcommand{\lan}{\langle}
\newcommand{\id}{\mathbb{I}}
\newcommand{\im}{\mathbbmss{1}}
\newcommand{\Cn}{\mathbb{C}}
\newcommand{\re}{\mathop{\mathbb{R}}\nolimits}
\newcommand{\esp}{\mathbb{E}}
\newcommand{\var}{\mathop{\mathrm{Var}} \nolimits}
\newcommand{\mse}{\mathop{\mathrm{MSE}} \nolimits}
\newcommand{\rlprt}{\mathop{\mathrm{Re}} \nolimits}

\newcommand{\imprt}{\mathop{\mathrm{Im}}\nolimits}
\newcommand{\tr}{\mathop{\mathrm{tr}}\nolimits}
\newcommand{\Tr}{\mathop{\mathrm{Tr}}\nolimits}

\newcommand{\smalloh}{\mathrm{o}}

\newcommand{\pr}{\prime}
\newcommand{\di}{\text{d}}
\newcommand{\trans}[1]{#1^{\top}}
\newcommand{\iu}{i}

\newcommand{\e}{\mathrm{e}}

\newtheorem{theorem}{Theorem}[section]

\newtheorem{lemma}[theorem]{Lemma}
\newtheorem{definition}[theorem]{Definition}
\newtheorem{remark}[theorem]{Remark}
\newtheorem{corollary}[theorem]{Corollary}
\newtheorem{claim}[theorem]{Claim}

\newcommand{\hil}{\mathcal{H}}

\newcommand{\su}{\mathfrak{su}}
\newcommand{\wt}{\widetilde}

\newcommand{\lil}{\mathcal{L}}

\newcommand{\sic}{\chi}
\newcommand{\mub}{\phi}
\newcommand{\td}{\tau}
\begin{document}
\bibliographystyle{h-physrev}
\title{Optimal estimation of $SU(d)$ using exact and approximate $2$-designs}
\author{Manuel A. Ballester}
\email{Manuel.Ballester@cwi.nl}
\homepage{http://homepages.cwi.nl/~balleste/}
\affiliation{Centrum voor Wiskunde en Informatica (CWI),
Kruislaan 413, \\P.O. Box 94079,
1098 SJ Amsterdam, The Netherlands}

\begin{abstract}
We consider the problem of estimating  an $SU(d)$ quantum operation when $n$ copies of it are available at the same time. It is well known that, if one uses a separable state as the input for the unitaries, the optimal mean square error will decrease as $1/n$. However it is shown here that, if a proper entangled state is used, the optimal mean square error will decrease at a $1/n^2$ rate. It is also shown that spherical $2$-designs (e.g.\ complete sets of mutually unbiased bases and symmetric informationally complete positive operator valued measures) can be used to design optimal input states. Although $2$-designs are believed to exist for every dimension, this has not yet been proven. Therefore, we give an alternative input state based on \emph{approximate} $2$-designs which can be made arbitrarily close to optimal. It is shown that measurement strategies which are based on local operations and classical communication between the ancilla and the rest of the system can be optimal.
\end{abstract}
\maketitle
\section{Introduction}
The problem of estimating a completely unknown $U\in SU(d)$ unitary operation is studied in this paper. It is assumed that $n$ copies of $U$ are available. The idea is to prepare a \emph{suitable} input state, use it as an input for $U^{\otimes n}$ and measure the output. One could also allow for an ancilla, i.e., a part of the input state that is left untouched. In addition to being interesting in itself, $SU(d)$ estimation also has applications in the problem of optimal alignment of reference frames  \cite{PeresScudo01a,Baganetal04b,Baganetal04c,Chiribellaetal04a}.

This problem has been considered from a Bayesian point of view for $SU(2)$ in Refs.\ \cite{Baganetal04b,Baganetal04c,Chiribellaetal04a,Hayashi06a} and for a general $SU(d)$ in Refs.\ \cite{Chiribellaetal05a,Kahn06a:qph}. They study the case where each copy of $U$ is used only once and obtain that the optimal mean square error (MSE) goes to zero at the rate $1/n^2$ compared to the $1/n$ rate that would be obtained if no entanglement in the input state were allowed.  Even though this problem is very interesting from a theoretical point of view, it is more likely that one does not have an arbitrary number of copies of a unitary gate, or that it is not yet practically feasible to create such a large entangled input state. Therefore, it would be more natural to assume that the number of copies $n$ is fixed and to repeat the experiment a number $N$ of times. Clearly the $N$-dependence of the MSE will be of the from $1/N$ as it is the case in models of the form $\rho^{\otimes N}$, so most of the effort will be on optimizing the $n$-dependence. Still one could compare the results obtained in this approach with the results obtained in \cite{Baganetal04b, Baganetal04c,Chiribellaetal04a,Hayashi06a,Kahn06a:qph}.  

The authors of  Ref.\ \cite{jietal06a:qph}, address the question of finding conditions for a general quantum operation to exhibit  this behavior.

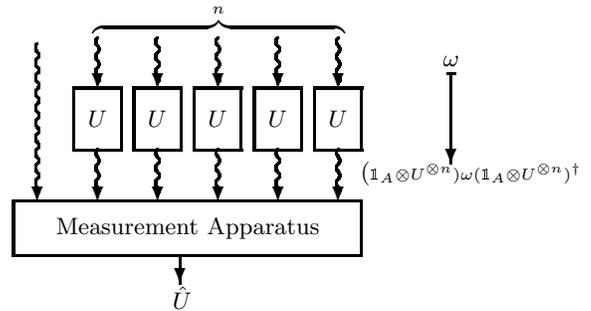
\begin{figure}[ht]
\begin{center}
\setlength{\unitlength}{0.20cm}
\begin{picture}(38,21)(0,0)
\thicklines
\multiput(6,12)(4,0){5}{\framebox(3,4){$U$}}
\thinlines 
\thicklines
\put(3.5,9){
\multiput(0,0)(0,1){10}{\qbezier(0,0)(0.25,0.25)(0,0.5)\qbezier(0,0.5)(-0.25,0.75)(0,1)}
}
\put(7.2,19.5){$\overbrace{\hspace{95pt}}^n$}

\put(30.5,17.5){$\omega$}
\put(30.7,17){\line(1,0){0.6}}
\put(31,17){\vector(0,-1){6}}
\put(25,10){($\scriptstyle \im_A\otimes U^{\otimes n})\omega(\im_A\otimes U^{\otimes n})^\dagger$}
\multiput(7.5,16.4)(4,0){5}
{
\multiput(0,0)(0,1){3}{\qbezier(0,0)(0.25,0.25)(0,0.5)\qbezier(0,0.5)(-0.25,0.75)(0,1)}
}
\multiput(7.5,16)(4,0){5}{\vector(0,-1){0}}
\multiput(7.5,9)(4,0){5}
{
\multiput(0,0)(0,1){3}{\qbezier(0,0)(0.25,0.25)(0,0.5)\qbezier(0,0.5)(-0.25,0.75)(0,1)}
}
\multiput(3.5,8.5)(4,0){6}{\vector(0,-1){0}}
\put(2,5){\framebox(23,3.5){Measurement Apparatus}}
\put(13,5){\vector(0,-1){2}}\put(12.5,1.4){$\hat{U}$}
\end{picture}
\end{center}
\caption{$n$ copies of $U$ are available at the same time.}
\label{fig:ncopiesofu}
\end{figure}

To state the problem more precisely, let $\omega$ on $\Cn^{d_A}\otimes\Cn^{d^n}$  be the input state, where $d_A$ is the dimension of the ancilla. The output state $\rho$, also on $\Cn^{d_A}\otimes(\Cn^{d})^{\otimes n}$, then becomes
\benn
\rho=(\im_A\otimes U^{\otimes n})\omega(\im_A\otimes U^{\otimes n})^\dagger,
\eenn
where $\im_A$ is the identity operator on the ancilla. This output state is then measured and the outcome of the measurement is recorded. This process is repeated $N$ times and from the measurement outcomes an \emph{estimate} of $U$ is made. The situation is represented in Fig.\ \ref{fig:ncopiesofu}.

In this paper, it is assumed that $U$ is parametrized by $\theta \in \re^{d^2-1}$, and an estimate $\hat{\theta}$ of $\theta$ is made. These considerations allow one to formulate the model studied in this paper as
\be\label{eq:model}
\rho_n(\theta,\omega)^{[N]}=\left([\im_A\otimes U(\theta)^{\otimes n}]\omega[\im_A\otimes U(\theta)^{\otimes n}]^\dagger\right )^{\otimes N}.
\ee

Obviously, if the input state $\omega$ is separable the situation would basically be the same as having only one copy of $U$ and repeating the experiment $nN$ times. In that case, results of classical statistics imply that the MSE would behave as $1/(nN)$. The question is whether by using an entangled state $\omega$ the dependence on $n$ can be improved. As one might expect, the answer to this question is positive.

In the rest of this paper it will be shown that, for $SU(d)$, there is  an input state, a measurement and an estimator, such that their MSE  vanishes at a $1/(Nn^2)$ rate. 
We use the so called \emph{pointwise approach}, which differs from the (more commonly used) Bayesian approach. In the Bayesian approach, both the measurement strategy
and the estimator are so chosen as to minimize the average of a loss function (often one minus fidelity) with respect to a given prior distribution for any $N$. 
In contrast, in the pointwise approach  one's goal is to optimize the performance of a scheme at a fixed point (the truth) in parameter space for large $N$.

In section \ref{sec:prelim} some quantum statistical results and notation are introduced. The notion of spherical $2$-designs, which will be useful for obtaining an optimal input state, is also defined. The figure of merit for evaluating  the performance of the input state and the measurement used, is also specified. In section \ref{sec:inputstate}, the optimal input state for our figure of merit is found. The state is optimal in the sense that it minimizes a lower bound on the MSE of any measurement (the quantum Cram\'er-Rao bound) and also in the sense that it guarantees the existence of a measurement that achieves this bound. It is shown that optimal input states can be constructed from $2$-designs (if they exist for dimension $d$). It is possible, however, that a construction of a $2$-design is not known in $\Cn^d$ (e.g.\ because they do not exist). We show in section \ref{sec:appr2d} that one can be arbitrarily close to optimal by using an approximate version of a $2$-design. In section \ref{sec:measurement} it is shown that there exists an optimal measurement strategy which can be performed by local operations and classical communication (LOCC) between the ancilla and the rest of the system.
Unfortunately, the optimal measurements shown here have the problem of being a different one for different values of $\theta$. This problem can be overcome in a two-step adaptive strategy like the one used in Ref.\ \cite{GillMassar00a}. Alternatively one could use the so-called random measurement. This measurement can be described as follows: at every repetition of the experiment, one  chooses a basis uniformly at random and measures in this basis. The random measurement gives the same performance regardless of what the actual $\theta$ is, but is only half as good as the optimal one, this is proven in the appendix.
\section{Preliminaries}\label{sec:prelim}
In this section some notions and results needed for the rest of the paper will be introduced. 
\subsection{Quantum statistics} \label{sec:qstats}
Let $\{M_\xi: \xi\in\Omega\}$, be a positive operator valued measure (POVM), where $\Omega$ is the outcome space. Let $\hat{\theta}_{\xi}$ be an unbiased estimator\footnote{Unbiased means that the expectation of the estimator is equal to the truth, i.e.,
$\esp \hat{\theta}_\alpha=\sum_{\xi\in\Omega}\tr[\rho(\theta) M_\xi]\hat{\theta}_{\xi\alpha}=\theta_\alpha.$} for $\theta \in \Theta \subset \re^{d^2-1}$, the $d^2-1$-dimensional parameter of a density matrix $\rho$. The MSE can then be written as the $(d^2-1)\times (d^2-1)$ matrix with elements
\benn
&V_n(\hat{\theta},\theta,M,\omega)^{[N]}_{\alpha \beta}\\
&=\sum_{\xi\in\Omega}\tr[\rho_n(\theta,\omega)^{[N]} M_\xi](\hat{\theta}_{\xi\alpha} -\theta_\alpha)(\hat{\theta}_{\xi\beta} -\theta_\beta).
\eenn
The Fisher information (FI) can be calculated as the $(d^2-1) \times (d^2-1)$ matrix with elements
\benn
&I_n(\theta,M,\omega)^{[N]}_{\alpha \beta}\\
&= \sum_{\xi\in \Omega_+} \frac{\tr[\rho_n(\theta,\omega)^{[N]}_{,\alpha} M_\xi] \tr[\rho_n(\theta,\omega)^{[N]}_{,\beta} M_\xi] }{\tr[\rho_n(\theta,\omega)^{[N]} M_\xi]},
\eenn
where  $f(\theta)_{,\alpha}$ means partial derivative of $f(\theta)$ with respect to $\theta_\alpha$ (later one the notation $\partial_\alpha f(\theta)$ will also be used), and $\Omega_+=\{\xi \in \Omega: \tr[\rho_n(\theta,\omega)^{[N]} M_\xi]>0\}$. The Cr\'{a}mer-Rao bound (CRB) \cite{BickelDoksum:book} states that
\benn
V_n(\hat{\theta},\theta,M,\omega)^{[N]}\geq [I_n(\theta,M,\omega)^{[N]}]^{-1}.
\eenn
The previous equation should be interpreted as a matrix inequality, i.e., $(V-I^{-1})$ is a positive semidefinite matrix.

If one performs a measurement $M$ which consists on repeating the same measurement $m$ on each of the copies then the measurement results will be independent and identically distributed, (i.i.d.)  and the Fisher information will satisfy 
\be \label{eq:FIiid}
I_n(\theta,M,\omega)^{[N]}=N I_n(\theta,m,\omega)^{[1]},
\ee
and it follows that
\benn
V_n(\hat{\theta},\theta,M,\omega)^{[N]}\geq \frac{[I_n(\theta,m,\omega)^{[1]}]^{-1}}{N}.
\eenn
It is a well known fact in mathematical statistics that (under some regularity conditions) the maximum likelihood estimator (MLE) in the limit of large $N$ is asymptotically unbiased and saturates the Cram\'er-Rao bound \cite{BickelDoksum:book}. Moreover no other reasonable estimator (unbiased or not) can do better \cite{GillLevit95a}. This means that it is enough to concentrate on finding a measurement that optimizes the Fisher information for $N=1$ and use the MLE.

The quantum Fisher information (QFI) is defined as the matrix with elements
\benn
H_n(\theta,\omega)^{[N]}_{\alpha \beta}=\tr[\rho_n(\theta,\omega)^{[N]} \lambda_n(\theta,\omega)^{[N]}_\alpha \circ \lambda_n(\theta,\omega)^{[N]}_\beta],
\eenn
where  $\{\lambda_n(\theta,\omega)^{[N]}_1,\ldots,\lambda_n(\theta,\omega)^{[N]}_p\}$ are the symmetric logarithmic derivatives (SLD), and  $A \circ B =(AB+BA)/2$. The SLD are defined as selfadjoint solutions to the equation
\benn
\partial_{\alpha}\rho_n(\theta,\omega)^{[N]} = \frac{\rho_n(\theta,\omega)^{[N]}\circ \lambda_n(\theta,\omega)^{[N]}_\alpha}{2}.
\eenn 
For pure state models the SLD can be chosen to be $\lambda_n(\theta,\omega)^{[N]}_\alpha=2\partial_{\alpha}\rho(\theta,\omega)^{(N,n)}$. From the definition of the QFI, it is easy to derive that
\benn
H_n(\theta,\omega)^{[N]}=NH_n(\theta,\omega)^{[1]},
\eenn
from now on $H_n(\theta,\omega)^{[1]}$ will simply be denoted by $H_n(\theta,\omega)$.

The Fisher information of any measurement is upper bounded by the QFI \cite{BraunsteinCaves94a}, i.e.,
\be \label{eq:bcii}
I_n(\theta,M,\omega)^{[N]}\leq H_n(\theta,\omega)^{[N]}=NH_n(\theta,\omega),~\forall M,
\ee
this is the Braunstein and Caves information inequality (BCII).
Furthermore if there is another real symmetric matrix $\wt{I}$ such that 
$\forall M,~I (\theta,M,\omega)^{(N,n)}\leq \wt{I},$
then it follows that $\wt{I}\geq H_n(\theta,\omega)^{[N]}$, i.e., the inequality (\ref{eq:bcii}) is sharp.  The BCII together with the CRB give rise to the quantum  Cr\'{a}mer-Rao bound (QCRB),
\be\label{eq:qcrb}
V_n(\hat{\theta},\theta,M,\omega)^{[N]} &\geq [H_n(\theta,\omega)^{[N]}]^{-1}\\
					&=\frac{[H_n(\theta,\omega)]^{-1}}{N}.
\ee 

The inequality given by Eq.\ (\ref{eq:bcii}) is in general not attainable: in general there is no measurement $M$ such that $I_n(\theta,M,\omega)^{[N]}= H_n(\theta,\omega)^{[N]}$. Because of this, it turns out that it is not always possible to compare the FI of different measurements. One has to choose what one wants to estimate by assigning weights to the different parameters, i.e., minimize an expression of the form $\Tr G V_n(\hat{\theta},\theta,M,\omega)^{[N]}$, where $G$ is a real positive semidefinite matrix, over all measurements, (reasonable) estimators and input states. Since the MLE asymptotically achieves equality in the CRB the problem can be reduced to minimizing $\Tr G [I(\hat{\theta},\theta,M,\omega)^{(N,n)}]^{-1}$ over all measurements, and input states. From Eqs.\ (\ref{eq:FIiid}) and  (\ref{eq:bcii}) it follows that the optimal FI  $I(\hat{\theta},\theta,M,\omega)^{(N,n)}\sim N$. It is therefore  meaningful to look at the quantity
\be \label{eq:figureofmerit}
C_n(\theta,\omega,G)=\lim_{N \to \infty} N \max_{M,\omega}\Tr G [I_n(\theta,M,\omega)^{[N]}]^{-1},
\ee
from (\ref{eq:bcii}) one readily obtains that 
\be\label{eq:Clowerbound}
C_n(\theta,\omega,G)\geq \Tr G [H_n(\theta,\omega)]^{-1}.
\ee

For pure state models (in our case when $\omega=|\Omega\ran \lan \Omega|$ for some $|\Omega\ran \in \Cn^{d_A}\otimes\Cn^{d^n}$), it has been shown \cite{Matsumoto02a} that (\ref{eq:bcii}) is attainable if and only if 
\be\label{eq:attbcii}
\imprt \tr\left [\rho_n(\theta,\omega)^{[1]} \lambda_n(\theta,\omega)^{[1]}_\alpha \lambda_n(\theta,\omega)^{[1]}_\beta\right]=0.
\ee
In that case the bound can be attained by independently performing the following measurement at each repetition
\be \label{eq:recattbcii}
m_{\xi}&=|m_{\xi}\ran \lan m_{\xi}|~~\xi\in\{1,\dots, d^21\},\\
m_{d^2+1}&=\im-\sum_{\xi=1}^{d^2} m_{\xi},
\ee
where
\benn
&|m_{\xi}\ran=\sum_{\chi=1}^{d^2} o_{\xi \chi} |b_{\chi}\ran,\\
&|b_\alpha\ran=\sum_{\beta=1}^{d^2-1} [H_n(\theta,\omega)]^{-\frac{1}{2}}_{\alpha \beta} \lambda_n(\theta,\omega)^{[1]}_\beta [\im_A\otimes U(\theta)^{\otimes n}]|\Omega\ran,\\
&|b_{d^2}\ran=[\im_A\otimes U(\theta)^{\otimes n}]|\Omega\ran,
\eenn
where $o$ a $d^2\times d^2$ real orthogonal matrix satisfying $o_{\xi,d^2}\neq 0$. 

The previous measurement has the drawback of depending on $\theta$, the actual value of the parameter, which is what one wants to estimate. This problem can be overcome by using a two step adaptive strategy like the one used in Ref.\ \cite{GillMassar00a}. One  spends $\sqrt{N}$ of the repetitions in finding a rough estimate $\hat{\theta}_1$ of $\theta$ using any informationally complete measurement. Then the measurement (\ref{eq:recattbcii}) is performed on the rest of the copies as if the truth were $\hat{\theta}_1$. This is also optimal \cite{HayashiMatsumoto03a:qph}.
\subsection{2-designs, MUBs and SIC-POVMs}

We will need the notion of mutually unbiased bases (MUBs),
which was introduced in~\cite{WoottersFields87a}. 
The following definition closely follows the one given in~\cite{Bandyopadhyayetal02a}.
\begin{definition}[MUBs] \label{def-mub}
Let $\mathcal{B}_1 = \{|\mub^1_1\ran,\ldots,|\mub^1_d\ran\}$ and $\mathcal{B}_2 =
\{|\mub^2_1\ran,\ldots,|\mub^2_d\ran\}$ be two orthonormal bases in
$\Cn^d$. They are said to be
\emph{mutually unbiased} if and only if
$|\lan\mub^1_i |\mub^2_j\ran| = 1/\sqrt{d}$, for every $i,j =
1,\ldots,d$. A set $\{\mathcal{B}_1,\ldots,\mathcal{B}_m\}$ of
orthonormal bases in $\Cn^d$ is called a \emph{set of mutually
unbiased bases} if each pair of bases is mutually unbiased.
\end{definition}
In any dimension $d$, the number of mutually unbiased bases is at
most $d+1$~\cite{Bandyopadhyayetal02a}. Explicit constructions are known if
$d$ is a prime power~\cite{Bandyopadhyayetal02a,WoottersFields87a}. Unfortunately not very much is known in other dimensions, for example,
it is still an open problem whether there exists a set of $7$ MUBs in dimension $d=6$.

The notion of symmetric informationally  complete POVMs (SIC-POVMs) \cite{Renesetal04a} will also be useful.
\begin{definition}[SIC-POVMs]
Let $\{|\sic_1\ran,\ldots,|\sic_{d^2}\ran\}$ be a set of state vectors in $\Cn^d$ satisfying $|\lan \sic_i|\sic_j\ran|=(d+1)^{-1}$ for every $i \neq j$. Then 
\benn
\left\{\frac{|\sic_1\ran \lan \sic_1|}{d},\ldots,\frac{|\sic_{d^2}\ran \lan \sic_{d^2}|}{d}\right\},
\eenn
is called a SIC-POVM.
\end{definition}
The fact that this actually is an informationally complete POVM follows from this definition \cite{Renesetal04a}.
They have been shown to exist for $d\in \{2,3,4,5,6,8\}$ and are conjectured to exist in all dimensions \cite{Zauner:thesis,Renesetal04a,Grassl04a:qph}.

It is easy to check that MUBs and SIC-POVMs satisfy the following property:
\benn
\frac{1}{d(d+1)}\sum_{b=1}^{d+1}\sum_{i=1}^{d} [|\mub^b_i\ran \lan \mub^b_i|]^{\otimes 2} &= 2\frac{\Pi_+^{(2,d)}}{d(d+1)},\\
\frac{1}{d^2}\sum_{i=1}^{d^2} [|\sic_i\ran \lan \sic_i|]^{\otimes 2}& = 2\frac{\Pi_+^{(2,d)}}{d(d+1)},
\eenn
where $\Pi_+^{(2,d)}$ is a projector onto the completely  symmetric subspace of $\Cn^d\otimes \Cn^d$. Indeed, one can straightforwardly check that the Hilbert-Schmidt distance between the right hand side and the left hand side is, in both cases, zero.

It was shown \cite{KlappeneckerRotteler05a} that any set of vectors satisfying this property, forms a \emph{spherical $2$-design}. 
More precisely, if a set of state vectors $\{|\td_1\ran,\ldots,|\td_{m}\ran\}$ in $\Cn^d$ form a $2$-design, then they satisfy
\be\label{eq:2ddef}
\frac{1}{m}\sum_{i=1}^m [|\td_i\ran \lan \td_i|]^{\otimes 2} = 2\frac{\Pi_+^{(2,d)}}{d(d+1)}.
\ee
For a formal definition of $2$-designs, see for example Refs.\  \cite{KlappeneckerRotteler05a,Renesetal04a}.
\subsection{A Chernoff bound for matrix valued random variables}
The following result, due to Ahlswede and Winter \cite{AhlswedeWinter02a} will be useful when dealing with approximate $2$-designs.
\begin{theorem}[Ahlswede and Winter]\label{thm:chernoffbound}
Let $X_1,\ldots,X_m$ be $p\times p$ i.i.d.\  selfadjoint random variables satisfying $0\leq X_b\leq \id$, $\esp X_b=M\geq\mu \id$ and $0\leq \epsilon\leq 1/2$. Then
\be\label{eq:chernoffbound}
\Pr\left[\left|\frac{1}{m}\sum_{b=1}^m X_b-M\right|>\epsilon M\right]\leq 2p\exp\left[-\frac{\epsilon^2\mu}{4\ln2}m\right].
\ee
\end{theorem}

\section{Optimal input state} \label{sec:inputstate}

From Eq.\ (\ref{eq:Clowerbound}) it is apparent that an optimal input state $\omega$ is one that minimizes $\Tr G [H_n(\theta,\omega)]^{-1}$. However it is possible that even if one minimizes this quantity, there is no measurement that achieves equality in (\ref{eq:Clowerbound}). Moreover the problem of minimizing $\Tr G [H_n(\theta,\omega)]^{-1}$ for a general $G$ is very hard, therefore we will concentrate in the case $G=\id$, the $(d^2-1) \times (d^2-1)$ identity matrix. With this choice of $G$, it will be shown that it is possible to minimize $\Tr[H_n(\theta,\omega)]^{-1}$ and at the same time, guarantee the existence of a measurement that achieves equality in Eq.\ (\ref{eq:Clowerbound}). 

Since from now on we work with $N=1$, we will write $(n)$ instead of $(1,n)$. Also the dependance on $\theta$ and $\omega$ will be omitted most of the times.

Since the QFI is convex \cite{Fujiwara01a}, the search for an optimal input state can be restricted to pure states. Let $|\Omega\ran\in \Cn^{d_A}\otimes\Cn^{d^n}$ be the input state, the output density matrix is
\benn
\rho^{(n)}=[\im_A \otimes U^{\otimes n}]|\Omega\ran\lan \Omega|[\im_A \otimes U^{\otimes n}]^{\dagger}.
\eenn

Let us define $\overline{\rho}_1$ as the average one-copy reduced density matrix of $\rho$, i.e.,
\benn
\overline{\rho}_1=\frac{1}{n}\sum_{s=1}^n \tr_{\bar{s}}\rho,
\eenn
where $\tr_{\bar{s}}$ means partial trace with respect to all copies except the $s^{th}$ one. In the same way, let us define $\overline{\rho}_2$ as the average symmetrized two-copy reduced density matrix of $\rho$, i.e.,
\benn
\overline{\rho}_2=\frac{1}{n(n-1)}\sum_{s\neq r}^n 
\frac{\tr_{\overline{s r}}\rho+W (\tr_{\overline{s r}}\rho) W}{2},
\eenn
where $\tr_{\overline{sr}}$ means partial trace with respect to all copies except the $r^{th}$ and the $s^{th}$, and $W$ is the swap operator $W: |\psi\ran|\phi\ran\mapsto|\phi\ran|\psi\ran$ 
for all $|\psi\ran,|\phi\ran \in \Cn^d$.  $W$ can be expressed as
$
W=\sum_{kl}|kl\ran\lan lk|
$
where $\{|k\ran\}$ is an orthonormal basis of $\Cn^d$.

\begin{lemma}
$H_n$ is given by
\be \label{eq:cqfi}
{H_n}_{\alpha \beta} =&4n\left( \rlprt \tr [\overline{{\omega}_B}_1 t_{\alpha}t_{\beta}]+{(n-1)}\tr\left[\overline{{\omega}_B}_2(t_\alpha \otimes t_\beta)\right]\right.\\
&-\left. n \tr[\overline{{\omega}_B}_1 t_{\alpha}] \tr[\overline{{\omega}_B}_1 t_{\beta}]\right),
\ee
where $t_\alpha = \iu U^{\dagger}U_{,\alpha}$ and ${\omega}_B=\tr_{\Cn^{d_A}}\omega$.
\end{lemma}
Note that $H_n$ depends only on $\overline{{\omega}_B}_2$ and that it will scale at most like $n^2$.
\begin{proof}
In this model the SLDs are
\benn
\lambda^{(n)}_\alpha = &2 [(\im_A \otimes \partial_\alpha U^{\otimes n})|\Omega\ran\lan \Omega|(\im_A \otimes U^{\otimes n})^{\dagger}\\
                                     &+(\im_A \otimes U^{\otimes n})|\Omega\ran\lan \Omega|(\im_A \otimes \partial_\alpha U^{\otimes n})]^{\dagger},
\eenn
so that
\be \label{eq:cqfiN}
L^{(n)}_{\alpha \beta}=&4[\lan \Omega|\im_A \otimes T^{(n)}_{\alpha}T^{(n)}_{\beta}|\Omega\ran \\
				&-\lan \Omega|\im_A \otimes T^{(n)}_{\alpha}|\Omega\ran\lan \Omega|\im_A \otimes T^{(n)}_{\beta}|\Omega\ran],
\ee
where $L^{(n)}_{\alpha \beta}=\tr \rho^{(n)} \lambda^{(n)}_\alpha \lambda^{(n)}_\beta$, 
\benn
T^{(n)}_{\alpha}& = \iu {U^{\otimes n}}^{\dagger}\partial_\alpha U^{\otimes n}\\
			 &= \sum_{s=1}^n\im^{\otimes(s-1)}\otimes t_\alpha \otimes \im^{\otimes(n-s)}\in \su(d^n),
\eenn
and $t_\alpha = \iu U^{\dagger}U_{,\alpha}\in \su(d)$. 

Let 
\benn
|\Omega\ran = \sum_{K=1}^{\min(d_A,d^n)} \sqrt{p_K}~|\psi^A_K\ran \otimes |\psi^B_K\ran,
\eenn
where $|\psi^A_K\ran $ ($|\psi^B_K\ran$) is a system of orthonormal vectors in $\Cn^{d_A}$ (respectively ${\Cn^d}^{\otimes n}$), then (\ref{eq:cqfiN}) may be rewritten as
\benn
L^{(n)}_{\alpha \beta} =4\left[\tr({\omega}_B T^{(n)}_{\alpha}T^{(n)}_{\beta})-\tr({\omega}_B T^{(n)}_{\alpha})\tr({\omega}_B T^{(n)}_{\beta})\right],
\eenn
where
\benn
{\omega}_B=\tr_{\Cn^{d_A}}|\Omega\ran\lan \Omega|=\sum_K p_K |\psi^B_K\ran\lan\psi^B_K|.
\eenn

Now, (\ref{eq:cqfiN}) may be rewritten as
\benn 
L^{(n)}_{\alpha \beta} =&4n( \tr [\overline{{\omega}_B}_1 t_{\alpha}t_{\beta}]+{(n-1)}\tr\left[\overline{{\omega}_B}_2(t_\alpha \otimes t_\beta)\right]\\
				&-n \tr[\overline{{\omega}_B}_1 t_{\alpha}] \tr[\overline{{\omega}_B}_1 t_{\beta}]),
\eenn
and the QFI is ${H_n}_{\alpha \beta}=\rlprt L^{(n)}_{\alpha \beta}$.
\end{proof}

The following lemma examines the conditions for which equality can be achieved in Eq.\ (\ref{eq:bcii}).

\begin{lemma}\label{lemma:attbcii}
There exists a measurement that achieves equality in the BCII (Eq.\ (\ref{eq:bcii})) if and only if 
\benn
\overline{{\omega}_B}_2=\frac{\im \otimes \im}{d^2} + \sum_{\alpha \beta} \wt{h}_{\alpha \beta} t_\alpha \otimes t_\beta,
\eenn
where $ \wt{h}_{\alpha \beta}= \wt{h}_{\beta \alpha}$.
\end{lemma}
\begin{proof}
Obviously, $\overline{{\omega}_B}_2$ is supported in the symmetric subspace of  $\Cn^d \otimes \Cn^d$. The most general state in the symmetric subspace can be written as
\be \label{eq:generalaverage2copystate}
\overline{{\omega}_B}_2=&\frac{\im \otimes \im}{d^2} +\sum_{\alpha}b_\alpha [\im\otimes t_\alpha+ t_\alpha \otimes \im]\\
					&+ \sum_{\alpha \beta} \wt{h}_{\alpha \beta} t_\alpha \otimes t_\beta,
\ee
where $\wt{h}_{\alpha\beta}=\wt{h}_{\beta\alpha}$, $\overline{{\omega}_B}_1$  is then 
\benn
\overline{{\omega}_B}_1=\frac{\im}{d} +d\sum_{\alpha}b_\alpha  t_\alpha.
\eenn
The condition (\ref{eq:attbcii}) reduces to 
\benn
\tr (\overline{{\omega}_B}_1 [t_{\alpha},t_{\beta}])=0, \forall \alpha,\beta.
\eenn
Since $\{t_1,\ldots,t_{d^2-1}\}$ span $\su(d)$, the Lie algebra of $SU(d)$, the previous equation implies that for any $r, s\in \su(d)$, $\tr (\overline{{\omega}_B}_1 [r,s])=0$. Furthermore, since any $t\in \su(d)$ can be written as the commutator of two other $\su(d)$ elements, we have that for all $t \in \su(d)$, $\tr (\overline{{\omega}_B}_1 t)=0$ which in turn implies that $\overline{{\omega}_B}_1=\im/d$ or $b_\alpha=0$. 
Therefore, $\overline{{\omega}_B}_2$ must be of the form
\benn 
\overline{{\omega}_B}_2=\frac{\im \otimes \im}{d^2} + \sum_{\alpha \beta} \wt{h}_{\alpha \beta} t_\alpha \otimes t_\beta.
\eenn
\end{proof}
From here on, the parametrization will be chosen in such a way that $\tr t_\alpha t_\beta=\delta_{\alpha \beta}$, this allows one to express $W$ as
\benn
W=\frac{\im \otimes \im}{d}+\sum_{\alpha=1}^{d^2-1}t_\alpha \otimes t_\alpha.
\eenn

The following lemma deals with minimizing $\Tr [H_n]^{-1}$. It turns out that the input states that minimize this quantity also satisfy the conditions of lemma \ref{lemma:attbcii} so that this minimum value can also be attained.

\begin{lemma} \label{lemma:lowerbound}
Any input state $\omega$ satisfies
\be\label{eq:hinvlowerbound}
C_n(\theta,\omega,\id)\geq \frac{d(d+1)^2(d-1)}{4n(n+d)},
\ee
with equality if and only if 
\be \label{eq:optimalaverage2copystate}
\overline{{\omega}_B}_2&=\frac{\im \otimes \im}{d^2} + \frac{1}{d(d+1)} \sum_{\alpha}  t_\alpha \otimes t_\alpha\\
				     &=\frac{1}{d(d+1)}(\im \otimes \im + W)=\frac{2}{d(d+1)}\Pi_+^{(2,d)}.
\ee 
\end{lemma}
\begin{proof}
The trace of $H_n$ for the most general symmetric $\overline{{\omega}_B}_2$ on $\Cn^d \otimes \Cn^d$ (\ref{eq:generalaverage2copystate}) can be written as
\benn
\Tr H_n = 4n [\tr {{\omega}_B}_1 \sum_\alpha t_\alpha^2+(n-1)\Tr \wt{h}-n d^2 \sum_\alpha b_\alpha^2],
\eenn
where $\wt{h}=[\wt{h}_{\alpha \beta}]$. The operator $\sum_\alpha t_\alpha^2$ is a Casimir operator and therefore proportional to the identity, the proportionality factor can be found to be $(d^2-1)/d$ by taking the trace, then
\benn
\Tr H_n = 4n [\frac{d^2-1}{d}+(n-1)\Tr \wt{h}-n d^2 \sum_\alpha b_\alpha^2].
\eenn

The trace of $\wt{h}$ can be easily found
\benn
\Tr\wt{h}  &=\sum_{\alpha}\wt{h}_{\alpha \alpha}=\tr [\overline{{\omega}_B}_2 \sum_{\alpha} t_\alpha \otimes t_\alpha]\\
		&=\tr \overline{{\omega}_B}_2 W-\frac{1}{d}= 1-\frac{1}{d},
\eenn
where we have used that $\overline{{\omega}_B}_2$ is supported in the symmetric subspace of $\Cn^d\otimes \Cn^d$. Therefore the trace of $H_n$ satisfies
\benn
\Tr H_n &= 4n\left[\frac{d^2-1}{d} +(n-1)\Tr \wt{h}-d^2 \sum_\alpha b_\alpha^2\right]\\
		  &\leq 4n\left[\frac{d^2-1}{d} +(n-1)\frac{d-1}{d}\right]\\
		  &=4\frac{d-1}{d}n(n+d),
\eenn
with equality if and only if $b_\alpha=0$.
 Using the Cauchy-Schwarz inequality and the previous equation one gets that
\benn
(d^2-1)^2 &=\left[\Tr \left([H_n]^{-1/2}[H_n]^{1/2}\right)\right]^2\\
		&\leq \Tr [H_n] \Tr([H_n]^{-1})\\
		&\leq 4\frac{d-1}{d}n(n+d)C_n(\theta,\omega,\id),
\eenn
which implies (\ref{eq:hinvlowerbound}). Equality is attained if and only if $H_n$ is proportional to $[H_n]^{-1}$ and  $b_\alpha$=0. $H_n$ is proportional to $[H_n]^{-1}$ if and only if it is proportional to the identity which happens if and only if $\wt{h}$ is proportional to the identity. Therefore we have that the optimal $\wt{h}$ is
\benn
\wt{h}=\frac{1}{d(d+1)} \id,
\eenn
and the optimal $\overline{{\omega}_B}_2$ is given by (\ref{eq:optimalaverage2copystate}).
 The QFI corresponding to this state is
\be\label{eq:optimalqfi}
H_n=4 \frac{n(n+d)}{d(d+1)} \id.
\ee
\end{proof}

Next one needs to find an input state $|\Omega\ran$ such that (\ref{eq:optimalaverage2copystate}) holds.
In the dimension where $2$-designs exist, they can also be used to construct input states that satisfy (\ref{eq:optimalaverage2copystate}). Indeed it is easy to check that if the vectors $\{|\td_1\ran,\ldots,|\td_m\ran\}$ form a $2$-design then the state
\be \label{eq:inputstate2d}
|\Omega\ran=\frac{1}{\sqrt{m}}\sum_{i=1}^m  |i\ran \otimes |\td_i\ran^{\otimes n},
\ee
satisfies (\ref{eq:optimalaverage2copystate}). In particular, the dimension of the ancilla would be $d_A=d^2$ if a SIC-POVM is used, and $d_A=d(d+1)$ if a set of $d+1$ MUBs is used.

We now have an  input state (given by (\ref{eq:inputstate2d})) that satisfies (\ref{eq:optimalaverage2copystate}) and therefore is optimal in the sense of lemma \ref{lemma:lowerbound}. However, we still need to check that it satisfies one more condition: there should be a one to one correspondence between unitaries $U$ and output states $[\im_A\otimes U^{\otimes n}]|\Omega\ran$, this is proven in the following lemma.
\begin{lemma} \label{lemma:1to1param}
The input states given by   (\ref{eq:inputstate2d}) satisfies
\benn
|\lan \Omega| [\im_A\otimes U_1^{\otimes n}]^{\dagger}[\im_A\otimes U_2^{\otimes n}]|\Omega\ran|=1
\eenn
if and only if $U_1^{\dagger} U_2$ is proportional to the identity, i.e., they can only differ by a multiplicative phase.
\end{lemma}
\begin{proof}
Let $U=U_1^{\dagger} U_2$ be diagonalized as  
\benn
U=\sum_{k=1}^d \e^{\iu \eta_k} |u_k\ran \lan u_k|.
\eenn

For the input state (\ref{eq:inputstate2d}), we have that
\benn
|\lan \Omega| [\im_A\otimes U_1^{\otimes n}]^{\dagger}[\im_A\otimes U_2^{\otimes n}]|\Omega\ran|&=\frac{1}{m}\left|\sum_{i=1}^m \lan \td_i|U|\td_i\ran^n\right|\\
&\leq\frac{1}{m}\sum_{i=1}^m \left|\lan \td_i|U|\td_i\ran\right|^n\\
&\leq 1.
\eenn
One of the conditions for equality is that  $|\lan \td_i|U|\td_i\ran|=1$ for all $i\in\{1,\ldots,m\}$, i.e., $\lan \td_i|U|\td_i\ran=\e^{\iu \phi_i}$. We have then that
\benn
\e^{\iu \phi_i}=\lan \td_i|U|\td_i\ran=\sum_{k=1}^d \e^{\iu \eta_k} |\lan k|\td_i\ran|^2,
\eenn
which implies that for all $i\in \{1,\ldots,m\}$ and $k\in \{1,\ldots,d\}$ either $\lan k|\td_i\ran=0$ or $\e^{\iu \eta_k}=\e^{\iu \phi_i}$. Next we will prove that for every $k\neq l\in \{1,\ldots,d\}$  there exists an $i\in \{1,\ldots,m\}$ such that both $\lan k|\td_i\ran\neq0$ and $\lan l|\td_i\ran\neq0$, this would imply that $\e^{\iu \eta_k}=\e^{\iu \eta_l}=\e^{\iu \phi_i}$ which would finish the proof. Indeed, we have that
\benn
\lan kl|\frac{1}{m}\sum_{i=1}^m [|\td_i\ran \lan \td_i|]^{\otimes 2}|kl\ran = \lan kl|\frac{2}{d(d+1)}\Pi_+^{(2,d)}|kl\ran,
\eenn
or
\benn
\frac{1}{m}\sum_{i=1}^m |\lan k|\td_i\ran \lan l | \td_i\ran|^{2} =\frac{1}{d(d+1)},
\eenn
wich implies that there must exist at least one $i$ such that $ |\lan k|\td_i\ran \lan l | \td_i\ran|>0$.
\end{proof}
It is now possible to state the main theorem.
\begin{theorem}\label{theorem:main}
 The input state given by  (\ref{eq:inputstate2d}) satisfies
 \begin{enumerate}
 \item  \label{item1} The map $U \mapsto [\im_A\otimes U^{\otimes n}]|\Omega\ran$ from $SU(d)$ to $\Cn^{d_A}\otimes(\Cn^{d})^{\otimes n}$ is injective.
 \item \label{item2}\benn C_n(\theta,\omega,\id) =\frac{d(d+1)^2(d-1)}{4n(n+d)}, \eenn which is optimal.
 \end{enumerate}
\end{theorem}
\begin{proof}
Point \ref{item1} is proven in lemma \ref{lemma:1to1param}. Point \ref{item2} is proven by lemma \ref{lemma:lowerbound} and the fact that the input state given by (\ref{eq:inputstate2d}) satisfies   Eq.\ (\ref{eq:optimalaverage2copystate}).
\end{proof}
All the main ingredients for the optimal estimation of $n$ copies of  a $SU(d)$ quantum operation have been proven. As input state one can choose (\ref{eq:inputstate2d}) if they exist for dimension $d$. The used ancilla has dimension $d_A\sim d^2$.This input state is optimal as proven in theorem \ref{theorem:main}. The output states are measured using the recipe given by (\ref{eq:recattbcii}), data are collected and an estimate of the parameter is given by using the MLE. 

The case where there is no known construction of a $2$-design in $\Cn^d$ is dealt with next.
\section{Approximate 2-designs}\label{sec:appr2d}
Let $\{U_1,\ldots,U_m\}$ be an i.i.d.\ sequence of unitaries chosen uniformly at random from the Haar measure. Let
\be\label{eq:approxinput}
|\Omega\ran = \frac{1}{\sqrt{md}}\sum_{k=1}^d \sum_{b=1}^m |bk\ran\otimes\left[U_b|k\ran\right]^{\otimes n},
\ee
and $\omega=|\Omega\ran\lan \Omega|$. For this choice we have that
\benn
\overline{{\omega}_B}_2&= \frac{1}{md}\sum_{k=1}^d \sum_{b=1}^m [U_b|k\ran\lan k|U_b^\dagger]^{\otimes 2},\\
\overline{{\omega}_B}_1&= \frac{\im}{d},
\eenn
which ensures that equality can be achieved in  the BCII (Eq.\ (\ref{eq:bcii})). It is also easy to check that 
\benn
\esp~\overline{{\omega}_B}_2 &=\frac{2}{d(d+1)}\Pi_+^{(2,d)}\\
\var~\overline{{\omega}_B}_2 &= \frac{1}{md} \esp~\overline{{\omega}_B}_2 ,
\eenn
i.e.\ the larger $m$ is, the closer $\overline{{\omega}_B}_2$ will be to satisfying (\ref{eq:2ddef}). This is why we call them approximate $2$-designs.

Using (\ref{eq:cqfi}) one can calculate the QFI corresponding to  the input state (\ref{eq:approxinput})
\benn
H_n(U_1,\ldots,U_m)=\frac{1}{m}\sum_{b=1}^m h_n(U_b),
\eenn
where
\benn
h_{n}(U)_{\alpha\beta}=&\frac{4n}{d}\left[ \delta_{\alpha \beta}\right.\\
					&\left.+(n-1)\sum_{k=1}^d\lan k|U^\dagger t_\alpha U|k\ran\lan k|U^\dagger t_\beta U|k\ran\right].
\eenn
\begin{lemma}
Let $H^0_n$ be the optimal QFI  (\ref{eq:optimalqfi}), then if 
\benn
m\geq \frac{4(d+1)\ln2}{\epsilon^2} \ln \left[\frac{2(d^2-1)}{1-q}\right],
\eenn
we have that 
\be\label{eq:boundsonH}
(1-\epsilon)[H^0_n]^{-1}\leq H_n^{-1}\leq(1+\epsilon)[H^0_n]^{-1},
\ee
holds with probability at least $q$.
\end{lemma}
\begin{proof}
The strategy is to apply theorem \ref{thm:chernoffbound} to 
\benn
X_b=\frac{[H^0_n]^{-1/2}h_n(U_b)[H^0_n]^{-1/2}}{d+1},~b\in\{1,\ldots,m\}.
\eenn
\begin{claim}\label{thm:xlessthan1}
$X_b\leq \id$.
\end{claim}
\begin{proof}[Proof of claim \ref{thm:xlessthan1}]
This will be done by showing that $h_n(U)\leq (d+1)H^0_n$ for all $U$. Indeed, let 
$x\in \re^{d^2-1}$, be a unit vector, and $t=\sum_\alpha x_\alpha t_\alpha$, we have that
\benn
\trans{x}h_n(U)x&=\sum_{\alpha \beta}x_\alpha h_n(U)_{\alpha\beta}x_\beta\\
			&=\frac{4n}{d}\left[ 1+(n-1)\sum_{k=1}^d\lan k|U^\dagger t U|k\ran^2\right]\\
			&\leq \frac{4n}{d}\left[1+(n-1)\sum_{kl} |\lan k|U^\dagger t U|l\ran|^2\right]\\
			&=\frac{4n^2}{d}<(d+1)\frac{4n(n+d)}{d(d+1)}\\
			&=(d+1)\trans{x}H^0_nx.
\eenn
Where we use the fact that $\tr t^2=1$ and that $n<(n+d)$.
Since the above equation holds for any $x\in\re^{d^2-1}$ we have  $h_n(U)\leq (d+1)H^0_n$ as desired.
\end{proof}
Next, we need the expectation of $h_n(u)$.
\begin{claim}\label{thm:espx}
\benn
\esp X_b=\frac{\id}{d+1}.
\eenn
\end{claim}
\begin{proof}[Proof of claim \ref{thm:espx}]
It suffices to prove that $\esp h_n(U)=H^0_n$.
\benn
&\esp h_n(U)_{\alpha\beta}=\int h_n(U)_{\alpha \beta} \di U\\
					&=\frac{4n}{d}\left[ \delta_{\alpha \beta} \phantom{1}^{\displaystyle \phantom{1}^{\displaystyle \phantom{1}}} \right.\\
					&\left.+(n-1)\sum_{k=1}^d \tr\left( [t_\alpha \otimes t_\beta] \int[U|k\ran\lan k|U^\dagger]^{\otimes 2}\di U \right)\right]\\
					&=\frac{4n}{d}\left[ \delta_{\alpha \beta} +\frac{2(n-1)}{d(d+1)}\sum_{k=1}^d \tr\left( [t_\alpha \otimes t_\beta] \Pi_+^{(2,d)}\right)\right]\\
					&=\frac{4n}{d}\left[ 1+\frac{(n-1)}{d+1}\right]\delta_{\alpha \beta} \\
					&={H^0_n}_{\alpha\beta}.
\eenn
\end{proof}
We can now apply theorem \ref{thm:chernoffbound} with $p=d^2-1$ and $\mu=1/(d+1)$ to get
\benn
\Pr\left[\left|\frac{1}{m}\sum_{b=1}^m X_b-\frac{\id}{d+1}\right|>\epsilon \frac{\id}{d+1}\right]&\\
\leq 2(d^2-1)\exp&\left[-\frac{\epsilon^2m}{4(d+1)\ln2}\right],
\eenn
or in terms of $H_n$
\benn
\Pr\left[\left|H_n-H^0_n\right|>\epsilon H^0_n\right]&\\
\leq 2(d^2-1)&\exp\left[-\frac{\epsilon^2m}{4(d+1)\ln2}\right].
\eenn
\end{proof}
The statement of the lemma follows immediately from the previous equation.

\begin{corollary}
 If 
\benn
m >\frac{4(d+1)\ln2}{\epsilon^2} \ln \left[2(d^2-1)\right],
\eenn
then there exists a choice $\{U_1,\ldots,U_m\}$ such that (\ref{eq:boundsonH})
holds. This implies that using these  unitaries in the input state (\ref{eq:approxinput}),
\benn
C_n(\theta,\omega,\id) \leq (1+\epsilon)\frac{d(d+1)^2(d-1)}{4n(n+d)}.
\eenn
Comparing this upper bound with the lower bound from lemma \ref{lemma:lowerbound} one can see  that by choosing $m$ large enough, one can be arbitrarily close to optimality.
\end{corollary}

We also need to prove an equivalent of lemma \ref{lemma:1to1param}, i.e., that a state of the form (\ref{eq:approxinput}) also gives a one to one correspondence between input states and unitaries.
\begin{lemma}
A state of the form (\ref{eq:approxinput}) also gives a one to one correspondence between input states and unitaries with probability $1$.
\end{lemma}
\begin{proof}
 As in lemma \ref{lemma:1to1param} it suffices to show that $|\lan \Omega|(\im_A\otimes U^{\otimes n})|\Omega\ran|=1$ holds if and only if $U$ is proportional to the identity.
\benn
|\lan \Omega|(\im_A\otimes U^{\otimes n})|\Omega\ran|&=\frac{1}{md}\left|\sum_{kb}\lan k|U_b^\dagger UU_b|k\ran^n\right|\\
						&\leq\frac{1}{md}\sum_{kb}|\lan k|U_b^\dagger UU_b|k\ran|^n\\
						&\leq 1,
\eenn
with equality  only if $|\lan k|U_b^\dagger UU_b|k\ran|=1$ for all $k$ and $b$. In particular, this means that for every $b$, $\{|k\ran\}$ is a basis of eigenvectors of $U_b^\dagger UU_b$, i.e.,
\benn
U_b^\dagger UU_b=\sum_{k=1}^d \e^{\iu \phi_{k}}|k\ran\lan k|,
\eenn
or 
\benn
U=\sum_{k=1}^d \e^{\iu \phi_{k}}U_b|k\ran\lan k|U_b^\dagger.
\eenn
Take now $b\neq b^\prime$, we have that
\be\label{eq:blah}
\e^{\iu \phi_k}=\lan k|U_b^\dagger UU_b|k\ran=\sum_{l=1}^d \e^{\iu \phi_l} |\lan k|U_b^\dagger U_{b^\prime}|l\ran|^2.
\ee
Since $U_b$ and $U_{b^\prime}$ where choosen unifornly at random, it is true that with probability $1$, $|\lan k|U_b^\dagger U_{b^\prime}|l\ran|^2>0$ for all  $l$. This, together with Eq.\ (\ref{eq:blah}), immediately implies that for all $l$, $\phi_l=\phi_k$ which in turn means that $U$ is proportional to the identity as desired.
\end{proof}

\section{Other measurement strategies (LOCC and random)}\label{sec:measurement}

\subsection{LOCC measurements}
It is interesting to see how the problem changes if the type of measurements that can be performed is restricted. Suppose, for example, that the measurement is performed by two parties, Alice and Bob. Suppose also, that Alice has access only to the ancilla and Bob only to the rest of the system.  If the input state is of the type (\ref{eq:inputstate2d}) Bob's reduced state is 
\benn
\rho_{B}=\frac{1}{m}\sum_{r=1}^m [U|\td_r\ran\lan \td_r|U^\dagger]^{\otimes n},
\eenn
a separable state. This means  that Bob's optimal estimation strategy will have an MSE which depends on $n$ as  $1/n$ at best. If Alice sends the ancilla to Bob, he will be able to achieve the  $1/n^2$ rate. In the case of MUBs and SIC-POVMs the ancilla will be small, its dimension  is of the order $d^2$, i.e., independent of $n$.

It is also interesting to ask what happens if Alice and Bob can exchange classical information. Consider the following simple LOCC measurement: Alice performs the measurement with components $A_r=|r\ran\lan r|$ on the ancilla and then sends the outcome to Bob. With this information Bob's state becomes
\benn
\rho_{B|r}= [U|\td_r\ran\lan \td_r|U^\dagger]^{\otimes n},
\eenn
which is also a product state so its Fisher information given $r$ will behave as $n$. The total Fisher information will be the average of the Fisher informations for fixed $r$ and so will also behave as $n$ and the MSE as $1/n$. Of course, this is a very special LOCC measurement, it turns out that, at least in the dimensions where  there exist $d+1$ MUBs, there exists an LOCC strategy which is optimal.
\begin{lemma}\label{loccstrategy}
If there exists a set of $d+1$ MUBs, then the bound (\ref{eq:hinvlowerbound}) can be attained using an LOCC measurement.
\end{lemma}

\begin{proof}
The lemma is proven by showing such a strategy. 

The output state is
\benn
|\psi\ran=\frac{1}{\sqrt{d(d+1)}}\sum_{b=1}^{d+1}\sum_{k=1}^d |bk\ran \otimes [U|\mub^b_k\ran]^{\otimes n}.
\eenn
 Alice measures performs in the ancilla the measurement with elements $A_{bk}=|b\ran\lan b|\otimes|f_k\ran\lan f_k|$, where 
\benn
|f_k\ran=\frac{1}{\sqrt{d}}\sum_{l=1}^d \exp\left[{\frac{2 \pi \iu k l}{d}}\right] |l\ran,
\eenn
is the Fourier transform of the basis $\{|k\ran\}$. She obtains outcomes $b,k$ with probability $[d(d+1)]^{-1}$ and communicates her outcome to Bob. In that case, Bob's state becomes
\be\label{eq:bobstate}
|\psi\ran_{B|bk}=\frac{1}{\sqrt{d}}\sum_{l=1}^d  \exp\left[{-\frac{2 \pi \iu k l}{d}}\right] [U|\mub^b_l\ran]^{\otimes n}.
\ee
He should still perform a measurement on this state. The Fisher information of this procedure is
\benn
I=\frac{1}{d(d+1)}\sum_{bk}I^{[bk]},
\eenn
where $I^{[bk]}$ is the Fisher information of Bob's measurement on the state (\ref{eq:bobstate}). The QFI for  the state (\ref{eq:bobstate}) can be calculated using Eq.\ (\ref{eq:cqfi})
\benn
H^{[bk]}_{\alpha \beta}=\frac{4n}{d}\left[\delta_{\alpha \beta}+(n-1)\sum_{l}\lan \mub^b_l|t_\alpha|\mub^b_l\ran \lan\mub^b_l|t_\beta|\mub^b_l\ran\right].
\eenn
Furthermore, the condition (\ref{eq:attbcii}) is satisfied therefore there exists a measurement (e.g.\ the measurement given by (\ref{eq:recattbcii})) which achieves equality between the QFI and the FI. Using the fact that the $d+1$ MUBs from a $2$-design, one can check that
\benn
I=\frac{1}{d(d+1)}\sum_{bk}H^{[bk]}=4\frac{n(n+d)}{d(d+1)},
\eenn
which is exactly the optimal value (\ref{eq:optimalqfi}).
\end{proof}
\begin{remark}
This proof can be easily adapted to show that for an input state of the form (\ref{eq:approxinput}) there exists an LOCC measurement such that there is equality between the FI and the QFI corresponding to that state.
\end{remark}
This result appears to contradict Ref.\ \cite{Ballester04a} where it was shown that for $n=1$, optimal collective measurements were at least $2(d+1)/d$ times better than any LOCC measurement. However there is no contradiction. In contrast with the present work, in Ref.\ \cite{Ballester04a} the ancilla had dimension $d$ and the state used was a maximally entangled state. This was enough to obtain the optimal QFI and to guarantee the existence of a measurement that attains the QCRB. Once this input state is fixed one obtains the mentioned advantage of optimal collective measurements over LOCC ones. Here it is shown that by allowing a larger ancilla, one can still be optimal, and LOCC measurements can perform as well as collective ones.
\subsection{The random measurement}

The only perhaps not so desirable feature of the measurements strategies described so far, is that they make use of the recipe given by (\ref{eq:recattbcii}). This recipe gives a different measurement for different values of the parameter, one may need to use an adaptive strategy like the one described at the end of section \ref{sec:qstats}. This undesired feature can be easily avoided at the cost of being suboptimal by using the so-called random measurement. The random measurement can be described in the following way: at every repetition choose an orthonormal basis of $\Cn^{d_A}\otimes(\Cn^d)^{\otimes n}$ and measure on that basis. 
\begin{lemma}
The random measurement $M_r$ achieves a FI which is half the QFI, ie.,
\be\label{eq:FIofrandomemeas}
I_n(\theta,M_r,\omega)= \frac{H_n(\theta,\omega)}{2} =2\frac{n(n+d)}{d(d+1)} \id,
\ee
so that 
\benn
\Tr[I_n(\theta,M_r,\omega)]^{-1}=\frac{d(d+1)^2(d-1)}{2n(n+d)}.
\eenn
\end{lemma}
One would achieve the same if one modifies the LOCC strategy from lemma \ref{loccstrategy} such that Alice does the same but Bob performs the random measurement on his part of the system ($[\Cn^d]^{\otimes n}$).

A general proof of (\ref{eq:FIofrandomemeas}) will be given in the appendix.

\section{Discussion}
We have found an estimation strategy with an MSE that behaves like
\benn
V_n(\hat{\theta},\theta,M,\omega)^{[N]}= \frac{d(d+1)}{4n(n+d)N}\id + \smalloh(1/N).
\eenn
This gives us a hint on how to tackle the problem in which one is allowed to use every copy of $U$ only once and $n$ is not fixed.  The strategy would be to divide the $n$ copies into $n^\epsilon$ groups of $n^{1-\epsilon}$ copies each where $\epsilon$ is an arbitrary small but strictly positive real number. Then for each of the $n^\epsilon$ groups, one would perform the optimal strategy described above in this paper independently. With this procedure, one would expect the MSE to behave as
\benn
\mse(\hat{\theta},\theta,M)= \frac{d(d+1)}{4n^{2-\epsilon}}\id + \smalloh(1/n^{2-\epsilon}),
\eenn
i.e., as close to the $1/n^2$ rate as one wants. However, in this situation for each $n$ the model is a different one, so we would not be in the familiar i.i.d.\ case either.

\section{Conclusions}
We have considered the problem of estimating an $SU(d)$ operation when  a fixed number, $n$, of copies is available. By allowing entanglement in the input state, we have found an optimal estimation strategy where the MSE vanishes at a $1/(N n^2)$ rate, where $N$ is the (large) number of times the experiment is repeated. This is much better than the $1/(Nn)$ rate that one would obtain if no entanglement in the input state were allowed. We have shown that the optimal input states can be constructed from $2$-designs, if they exist for dimension $d$, otherwise from approximate versions of them. In both cases, these input states have another interesting property: if one has no access to the ancilla, the reduced state is separable and thus the MSE will behave as $1/(Nn)$ at best. We have also shown that, if a set of $d+1$ MUBs exists or if one uses an approximate $2$-design, classical information about the ancilla is actually enough to achieve equality in the Braunstein and Caves information inequality. In particular this means that in the former case one can be optimal with LOCC measurements. The optimal measurements found here are adaptive ones. It was shown that the so-called random measurement can be used to avoid this at the cost of being suboptimal. It would be interesting to find a non-adaptive measurement which is optimal for all values of the parameter.

\acknowledgments
I would like to thank Richard Gill, Madalin Gu\c{t}\u{a} and Masahito Hayashi for their very useful comments in the initial stage of this work. I have also benefited greatly from discussions with Stephanie Wehner.
This research was funded in its initial stage, by the Netherlands Organization for
Scientific Research (NWO), and the RESQ (IST-2001-37559)
project of the IST-FET programme of the European Union. Support of  the EU project QAP (IST-2005-15848) is also acknowledged.

\appendix
\section{The random measurement}

Suppose we have any pure state model on a $d$-dimensional Hilbert space $\hil$ (i.e. the number of parameters $p$ can be anything between $1$ and $2(d-1)$). It will be proven that the  random measurement $M_r$, which consists of choosing a basis of $\hil$ uniformly at random and then measuring on that basis, achieves
\be \label{eq:qfiofrandommeas}
I(\theta,M_r)=\frac{1}{2}H(\theta),
\ee
in particular, this measurement would be optimal when $p=2(d-1)$ if asymptotic fidelity is taken as the figure of merit. This measurement has also been studied in \cite{Hayashi98a}.

Let $|\psi\ran$  and $\{\lambda_1,\ldots,\lambda_p\}$, be the state vector and SLDs at the true value of the parameter respectively, and  let $\{|k\ran\}$ be any basis of $\hil$. It is an easy exercise to show that
\benn
I(\theta,\{|k\ran\lan k|\})_{\alpha \beta} = 	&\frac{1}{2}H(\theta)_{\alpha \beta} \\
								&+\frac{1}{2} \sum_{k=1}^d \rlprt \left[ \lan k|l_\alpha\ran \lan k|l_\beta\ran \frac{\lan \psi|k\ran}{\lan k|\psi\ran} \right],
\eenn
where $|l_\alpha\ran=\lambda_\alpha|\psi\ran$. Let $\lil$ be the subspace of $\hil$ spanned by $\{|l_1\ran,\ldots,|l_p\ran\}$ and $\mathcal{R}$ be its orthogonal complement, let $\im_{\lil}$  and $\im_{\mathcal{R}}$ be projectors onto $\lil$ and $\mathcal{R}$ respectively. Let the unitary operator $Y$ be defined as
\benn
Y= \im_\mathcal{R} +\iu \im_\lil.
\eenn
One should keep in mind that $V$  depends on the real value of the parameter, in particular, $Y|\psi\ran=|\psi\ran$ and $Y|l_\alpha\ran=\iu |l_\alpha\ran$. Clearly $\{Y|k\ran\}$ is a new basis and it is easy to see that
\benn
I(\theta,\{Y|k\ran\lan k|Y^\dagger\})_{\alpha \beta} &= \frac{1}{2}H(\theta)_{\alpha \beta} \\
										&-\frac{1}{2} \sum_{k=1}^d \rlprt \left[ \lan k|l_\alpha\ran \lan k|l_\beta\ran \frac{\lan \psi|k\ran}{\lan k|\psi\ran} \right].
\eenn
Let $M$ be the measurement with elements 
\benn
\left \{\frac{1}{2}|1\ran\lan 1|,\ldots,\frac{1}{2}|d\ran\lan d|, \frac{1}{2}Y|1\ran\lan 1|Y^\dagger,\ldots,\frac{1}{2}Y|d\ran\lan d|Y^\dagger \right\},
\eenn
its Fisher information at the truth is
\benn
I(\theta,M)&=\frac{1}{2}\left[I(\theta,\{|k\ran\lan k|\})+I(\theta,\{Y|k\ran\lan k|Y^\dagger\})\right]\\
		&=\frac{1}{2}H(\theta).
\eenn

Since $\{|k\ran\}$ is any basis, also the measurement $M_U$ with elements 
\benn
&\left \{\frac{1}{2}U|1\ran\lan 1|U^\dagger,\ldots,\frac{1}{2}U|d\ran\lan d|U^\dagger,\right.\\
&\left.\frac{1}{2}YU|1\ran\lan 1|U^\dagger Y^\dagger,\ldots,\frac{1}{2}YU|d\ran\lan d|U^\dagger Y^\dagger \right\},
\eenn
where $U$ is any unitary will also  satisfy
\benn
I(\theta,M_U)=\frac{1}{2}H(\theta).
\eenn

It is not hard to see that choosing $U$ at random and performing $M_U$ is the random measurement and therefore (\ref{eq:qfiofrandommeas}) must hold. This can also be shown analytically.  The Fisher information for the random measurement is
\benn 
I(\theta,M_r)	&=\int\mu(\di g) I(\theta,\{U_g|k\ran\lan k|U_g^\dagger\})\\ 
		&=\frac{1}{2}\int\mu(\di g) I(\theta,\{U_g|k\ran\lan k|U_g^\dagger\})\\
		&+\frac{1}{2}\int\mu(\di g^\pr) I(\theta,\{U_{g^\pr}|k\ran\lan k|U_{g^\pr}^\dagger\}),
\eenn
where $\mu$ is the normalized Haar measure and the integrals are over $SU(d)$.
Now let $g^\pr=hg$ and $U_h=Y$, we get
\benn
 I(\theta,M_r)=&\frac{1}{2}\int\mu(\di g) I(\{\theta,U_g|k\ran\lan k|U_g^\dagger\})\\
 			&+\frac{1}{2}\int\mu(\di g) I(\theta,\{YU_{g}|k\ran\lan k|U_{g}^\dagger Y^\dagger\})\\
		 =&\frac{1}{2}\int\mu(\di g) H(\theta)=\frac{1}{2}H(\theta),
\eenn
where we have used that 
\benn
I(\theta,\{U_g|k\ran\lan k|U_g^\dagger\})+I(\theta,\{YU_{g}|k\ran\lan k|U_{g}^\dagger Y^\dagger\})=H(\theta).
\eenn

In the case studied here, the random measurement would consist of choosing a basis of $\Cn^{d_A}\otimes (\Cn^{d})^{\otimes n}$ uniformly at random (with respect to the normalized Haar measure) and then measuring on that basis. Such a measurement is, as it was mentioned before, independent of the parameter and achieves
\benn
I_n(\theta,M_r,\omega)= \frac{H_n(\theta,\omega)}{2} =2\frac{n(n+d)}{d(d+1)} \id.
\eenn


\begin{thebibliography}{10}

\bibitem{PeresScudo01a}
A.~Peres and P.~F. Scudo,
\newblock Phys. Rev. Lett. {\bf 86}, 4160 (2001),
  \href{http://www.arxiv.org/abs/quant-ph/0010085/}{quant-ph/0010085}.

\bibitem{Baganetal04b}
E.~Bagan, M.~Baig, and R.~{Mu\~{n}oz-Tapia},
\newblock Phys. Rev. A {\bf 69}, 050303 (2004),
  \href{http://www.arxiv.org/abs/quant-ph/0303019/}{quant-ph/0303019}.

\bibitem{Baganetal04c}
E.~Bagan, M.~Baig, and R.~{Mu\~{n}oz-Tapia},
\newblock Phys. Rev. A {\bf 70}, 030301 (2004),
  \href{http://www.arxiv.org/abs/quant-ph/0405082/}{quant-ph/0405082}.

\bibitem{Chiribellaetal04a}
G.~Chiribella, G.~M. D'Ariano, P.~Perinotti, and M.~F. Sacchi,
\newblock Phys. Rev. Lett. {\bf 93}, 180503 (2004),
  \href{http://www.arxiv.org/abs/quant-ph/0405095/}{quant-ph/0405095}.

\bibitem{Hayashi06a}
M.~Hayashi,
\newblock Phys. Lett. A {\bf 354}, 183 (2006),
  \href{http://www.arxiv.org/abs/quant-ph/0407053/}{quant-ph/0407053}.

\bibitem{Chiribellaetal05a}
G.~Chiribella, G.~M. D'Ariano, and M.~F. Sacchi,
\newblock Phys. Rev. A {\bf 72}, 042338 (2005),
  \href{http://www.arxiv.org/abs/quant-ph/0506267/}{quant-ph/0506267}.

\bibitem{Kahn06a:qph}
J.~Kahn,
\newblock Preprint  (2006),
  \href{http://www.arxiv.org/abs/quant-ph/0603115/}{quant-ph/0603115}.

\bibitem{jietal06a:qph}
Z.~Ji, G.~Wang, R.~Duan, Y.~Feng, and M.~Ying,
\newblock Preprint  (2006),
  \href{http://www.arxiv.org/abs/quant-ph/0610060/}{quant-ph/0610060}.

\bibitem{GillMassar00a}
R.~D. Gill and S.~Massar,
\newblock Phys. Rev. A {\bf 61}, 042312 (2000),
  \href{http://www.arxiv.org/abs/quant-ph/9902063/}{quant-ph/9902063}.

\bibitem{BickelDoksum:book}
P.~J. Bickel and K.~A. Doksum,
\newblock {\em Mathematical Statistics. Basic Ideas and Selected Topics}
  (Prentice Hall, New Jersey, 2001).

\bibitem{GillLevit95a}
R.~D. Gill and B.~Y. Levit,
\newblock Bernouilli {\bf 1}, 59 (1995).

\bibitem{BraunsteinCaves94a}
S.~L. Braunstein and C.~M. Caves,
\newblock Phys. Rev. Lett. {\bf 72}, 3439 (1994).

\bibitem{Matsumoto02a}
K.~Matsumoto,
\newblock J. Phys. A: Math. Gen. {\bf 35}, 3111 (2002),
  \href{http://www.arxiv.org/abs/quant-ph/9711008/}{quant-ph/9711008}.

\bibitem{HayashiMatsumoto03a:qph}
M.~Hayashi and K.~Matsumoto,
\newblock Preprint  (2003),
  \href{http://www.arxiv.org/abs/quant-ph/0308150/}{quant-ph/0308150}.

\bibitem{WoottersFields87a}
W.~K. Wootters and B.~Fields,
\newblock Ann. Phys. {\bf 191} (1989).

\bibitem{Bandyopadhyayetal02a}
S.~Bandyopadhyay, P.~O. Boykin, V.~P. Roychowdhury, and F.~Vatan,
\newblock Algorithmica {\bf 34}, 512 (2002),
  \href{http://www.arxiv.org/abs/quant-ph/0103162/}{quant-ph/0103162}.

\bibitem{Renesetal04a}
J.~M. Renes, R.~Blume-Kohout, A.~J. Scott, and C.~M. Caves,
\newblock J. Math. Phys. {\bf 45}, 2171 (2004),
  \href{http://www.arxiv.org/abs/quant-ph/0310075/}{quant-ph/0310075}.

\bibitem{Zauner:thesis}
G.~Zauner,
\newblock {\em Quantendesigns-Grundz{\"u}ge einer nichtkommutativen
  Designtheorie},
\newblock PhD thesis, {Universit\"at} Wien, 1999.

\bibitem{Grassl04a:qph}
M.~Grassl,
\newblock Preprint  (2004),
  \href{http://www.arxiv.org/abs/quant-ph/0406175/}{quant-ph/0406175}.

\bibitem{KlappeneckerRotteler05a}
A.~Klappenecker and M.~R{\"{o}}tteler,
\newblock Mutually unbiased bases are complex projective $2$-designs,
\newblock in {\em Proc. Int. Symp. on Inf. Theory}, pp. 1740-- 1744, 2005,
  \href{http://www.arxiv.org/abs/quant-ph/0502031/}{quant-ph/0502031}.

\bibitem{AhlswedeWinter02a}
R.~Ahlswede and A.~Winter,
\newblock IEEE Trans. Inform. Theory {\bf 48}, 569 (2002),
  \href{http://www.arxiv.org/abs/quant-ph/0012127/}{quant-ph/0012127}.

\bibitem{Fujiwara01a}
A.~Fujiwara,
\newblock Phys. Rev. A {\bf 63}, 042304 (2001).

\bibitem{Ballester04a}
M.~A. Ballester,
\newblock Phys. Rev. A {\bf 69}, 022303 (2004),
  \href{http://www.arxiv.org/abs/quant-ph/0305104/}{quant-ph/0305104}.

\bibitem{Hayashi98a}
M.~Hayashi,
\newblock J. Phys. A: Math. Gen. {\bf 31}, 4633 (1998),
  \href{http://www.arxiv.org/abs/quant-ph/9704041/}{quant-ph/9704041}.

\end{thebibliography}
\end{document}